\documentclass[sigconf]{acmart}

\AtBeginDocument{%
  }

\setcopyright{acmlicensed}
\copyrightyear{2018}
\acmYear{2018}
\acmDOI{XXXXXXX.XXXXXXX}

\acmConference[Conference acronym 'XX]{Make sure to enter the correct
  conference title from your rights confirmation emai}{June 03--05,
  2018}{Woodstock, NY}
\acmISBN{978-1-4503-XXXX-X/18/06}

\begin{document}

\title{Building Bridges between Users and Content across Multiple Platforms during Natural Disasters}



\author{Lynnette Hui Xian Ng}
\affiliation{%
  \institution{Carnegie Mellon University}
  \city{Pittsburgh}
  \country{USA}}
\email{lynnetteng@cmu.edu}

\author{David Farr}
\affiliation{%
  \institution{University of Washington}
  \city{Seattle}
  \country{USA}
   \email{blah@blah.com}
}

\author{Iain Cruickshank}
\affiliation{%
 \institution{Carnegie Mellon University}
 \city{Pittsburgh}
 \country{USA}
 \email{icruicks@andrew.cmu.edu}
 }


\begin{abstract}
Social media is a primary medium for information diffusion during natural disasters. The social media ecosystem has been used to identify destruction, analyze opinions and organize aid. While the overall picture and aggregate trends may be important, a crucial part of the picture is the connections on these sites. These bridges are essential to facilitate information flow within the network. 
In this work, we perform a multi-platform analysis (X, Reddit, YouTube) of Hurricanes Helene and Milton, which occurred in quick session to each other in the US in late 2024. We construct network graphs to understand the properties of effective bridging content and users. We find that bridges tend to exist on X, that bridging content is complex, and that bridging users have relatable affiliations related to gender, race and job. Public organizations can use these characteristics to manage their social media personas during natural disasters more effectively.
\end{abstract}



\maketitle

\section{Introduction}

Social media has become a critical tool during natural disasters, serving as a platform for authorities and organizations to provide public information, acquire situational awareness, and coordinate grassroots aid \cite{zhang2019social}. Simultaneously, the distributed and participatory nature of social media enables the collection of valuable, localized information through everyday digital traces and interactions of users \cite{cresci2015linguistically}. This crowd-sourced data has proven useful for tasks such as damage assessment \cite{cresci2015linguistically}, constructing situational awareness overviews \cite{rudra2018extracting}, and modeling information diffusion \cite{dong2018information}.

In addition to aggregate narrative trends and identifying influential users, another critical but often overlooked aspect of social media is the \textbf{connections} within the network. These connections, referred to as \textit{bridges}, facilitate the flow of information across different communities and are vital during natural disasters. Bridges ensure that key insights reach diverse social media groups, helping to disseminate life-saving information effectively \cite{liu2019identification}.

Bridges manifest in multiple forms:

\begin{enumerate}
    \item \textbf{Intra-platform bridges:} Within a single social media platform, bridges can:
    \begin{itemize}
        \item Represent \textit{topic-oriented communities} connected through shared narratives or discussions \cite{carley2020social}.
        \item Be users who hold structural importance in the network, such that their removal disrupts clusters of nodes \cite{rabchevskiy2021method}.
        \item Include not only human users but also inauthentic bots, which can artificially link highly connected user clusters to influence information flow \cite{ng2023deflating}.
    \end{itemize}

    \item \textbf{Cross-platform bridges:} Bridges also operate across different social media platforms. While platforms have distinct ecosystems and rules, narratives about significant events often propagate across them \cite{ng2022cross}. Examples include:
    \begin{itemize}
        \item \textit{Linker posts}, where users on one platform (e.g., X) direct audiences to content on another (e.g., YouTube).
        \item \textit{Cross-platform bridging users}, who consistently amplify content from specific users across multiple platforms, boosting visibility and coherence in messaging \cite{murdock2023identifying}.
    \end{itemize}
\end{enumerate}

Bridges, whether within or across platforms, play a critical role in the complex social media ecosystem during natural disasters. Understanding and identifying these bridges is essential to enhancing information dissemination, mitigating misinformation, and improving disaster response effectiveness.

In late 2024, two hurricanes happened in the US within a week of each other, sparking discussions on social media that highlight the use and importance of bridges on these sites. Hurricane Helene was a Category 4 hurricane that affected the Carolinas, Georgia and the surrounding areas. Hurricane Milton was a Category 5 hurricane that sent orders of evacuations, and mostly affected Florida and Georgia. We collected data from X, YouTube and Reddit that relates to the discourse on these hurricanes, and analyzed the properties of bridging narratives and users. Using our analysis, we establish general characteristics of social media bridges, and elaborate on how such bridges can be leveraged upon for disaster management.


\paragraph{\textbf{Research Questions}}
In this work, we use a combination of linguistic analysis and network science to distill the features of content and user bridges within the hurricane discourse. Using our results, we establish parameters for building bridges within social media discourse. We ask one over-arching research question: \textbf{What makes an effective bridge?} To answer this question, we formulate three sub-research questions:  
\begin{itemize}
    \item What platforms build bridges?
    \item What types of content build bridges?
    \item What types of users build bridges?
\end{itemize}

\section{Methodology}
\subsection{Data}
We use the InformationTracer\footnote{\url{https://informationtracer.com}} platform to collect data. We use the search terms ``hurricane helene", and ``hurricane milton". We retrieve data from three social media platforms: X, YouTube and Reddit. This process retrieved tweets from X, video captions from YouTube, and original posts from Reddit. \autoref{tab:datastats} presents the statistics of the data. The data was collected on the week of each of the hurricanes: 24-29 September 2024 for Helene and 5-12 October 2024 for Milton.

\begin{table}
    \centering
    \begin{tabular}{ccccc}
        \toprule
        ~ & \multicolumn{2}{c}{\textbf{Helene}} & \multicolumn{2}{c}{\textbf{Milton}} \\ \midrule
        \textbf{Platform} & \textbf{Posts} & \textbf{Users} & \textbf{Posts} & \textbf{Users} \\ \midrule
        \textbf{X} & 5000 & 3613 & 5000 & 4054 \\
        \textbf{YouTube} & 500 & 225 & 500 & 114 \\
        \textbf{Reddit} & 247 & 124 & 239 & 272 \\ 
         \bottomrule
    \end{tabular}
    \caption{Statistics of Data Collected}
    \label{tab:datastats}
\end{table}

\subsection{Data Preprocessing}
The texts collected are first preprocessed to remove stopwords and social media artifacts like hashtags, @-mention symbols, etc. Retweets are kept because they signify a repetition of a particular message. The texts are then vectorized using a standard standard BERT-based embedding model\footnote{\url{https://huggingface.co/sentence-transformers/all-mpnet-base-v2}}. This model is chosen for its ability to vectorize both sentences and paragraphs into the same vector space as we have short texts from X, and varying lengths of texts from YouTube and Reddit.

\subsection{Identifying Bridging Nodes}
In graph theory, a bridge is a link whose deletion increases the number of connected components in a graph \cite{szell2010measuring,axler2000graduate}. We adapt this idea to bridging nodes, and define a bridging node as a node whose removal and the removal of its links increases the shortest path between its 1-degree connected nodes, and increases the number of connected components \cite{mediumIntroductionGraph}.

We modeled the spread of narratives through two network graphs: the \texttt{Content} graph and the \texttt{Users} graph. The \texttt{Content} graph is constructed by comparing the similarity of texts from posts. The nodes represent texts, and links represent similarity of the texts. Text similarity is calculated in the following manner: The vectors of preprocessed texts are compared with an all-pairs cosine comparison. Two texts are deemed similar if their cosine similarity is greater than 0.8, which is a threshold used in other text network constructions \cite{de2015learning,ng2022cross}. Bridges in the \texttt{Content} graph show how narratives are interconnected.

The \texttt{Users} graph is constructed from users that have similar texts. This uses the matrix transformation procedure defined in
\cite{ng2022cross} to identify users that are similar through their textual content in a cross-platform setting. In this graph, the nodes are users. Two users have a link if they have similar content. Bridges in the \texttt{Users} graph show which users connect over macro-textual elements like narratives.

After constructing both graphs, we apply the Leiden clustering algorithm on the graphs\cite{traag2019louvain}. 
We pruned the graph to retain only the top clusters; that is, clusters that have more than 10 nodes.
After pruning the graph, the \texttt{Content} network is left with 9 Leiden groups consisting of a total of 10,138 texts. The \texttt{User} network is left with 7 Leiden groups consisting a total of 789 users. From these pruned graphs, we find bridges using the Network Chain Decomposition method \cite{schmidt2013simple} from the NetworkX library\footnote{\url{https://networkx.org}}. This method decomposes the graph into connected chains of nodes, then calculate betweenness centrality of the nodes. Nodes with high betweenness centrality are potential bridging nodes, because they lie on the shortest paths of multiple nodes. The algorithm then disrupts the chain by removing these suspected bridging nodes and checks if the graph becomes more disconnected (i.e. increase in the number of connected components).

\subsection{Analysis of Bridging Nodes}
\paragraph{\textbf{Calculating Multi-platform linkages}} We analyzed the proportion of bridging nodes against the number of nodes of each platform. Then, we calculated the multi-platform linkages between the nodes. We analyzed the links for both \texttt{Content} and \texttt{Network} graphs in terms of platform pairs of the nodes that they link between. We plot \texttt{Content-Platform} and \texttt{Content-User} graphs, where nodes are social media platforms, and links are the extent to which the nodes are linked together in the original graph through similarity of \texttt{Content} or \texttt{User} respectively. For the \texttt{User} graph, we excluded self-loops, where the user posted similar text to himself.
We also calculated the proportion of cross-platform linkers, which are nodes (\texttt{Content} and \texttt{User} respectively) that bridge between two platforms.

\paragraph{\textbf{Calculating network centrality metrics}} For both the \texttt{Content} and \texttt{User} networks, we calculate four centrality metrics of bridging and non-bridging users. These metrics are: (1) total degree centrality, which measures the number of incident edges to a node. Nodes with many connections have high degree centrality, and can act as connectors between graph components. (2) Eigenvector centrality, which is the measure of how connected the node is to other influential nodes. (3) Hub centrality, which measures the extent the node links to authority or influential nodes. Nodes with high eigenvector or hub centrality can be effective information disseminators because they can spread information to other influential nodes. (4) Betweenness centrality, which is the extent to which a node lies on the shortest path of other pairs of nodes. Nodes with high betweenness centrality lie on the shortest path of many other sets of nodes.

\paragraph{\textbf{Calculating linguistic cues}}
To calculate linguistic cues, we use the NetMapper software\footnote{\url{https://netanomics.com/netmapper/}}. This software identifies psycholinguistic cues from the texts using a lexicon and has been used in many social media studies. We extracted linguistic cues for the texts and the usernames. The salient linguistic cues are picked out and compared between bridging and non-bridging nodes.

\paragraph{\textbf{Topic analysis of content}} For each Leiden cluster in the \texttt{Content} network, we perform Latent Dirichlet Allocation on the vectors of the preprocessed texts. For each cluster, we retrieve the words associated with the top 3 topics. The resultant topics are manually interpreted and annotated in \autoref{fig:metrics}(b).

\paragraph{\textbf{Bot user annotation}} To understand whether bridging nodes occur organically or inorganically, we annotate the users as bot or human using BotBuster \cite{ng2023botbuster}. BotBuster uses a mixture-of-experts architecture, where each expert is a machine learning model to evaluate each subset of data. This algorithm is ideal for our dataset because the common data pillar for each of the three platforms is the user name. Therefore, we only activate the user name expert when using BotBuster. The algorithm returns $P(bot)\in$ [0,1]. As per the original paper, we used the threshold of 0.5; therefore, $P(bot)\geq0.5 = bot$ and $P(bot)<0.5 = human$. 

\paragraph{\textbf{User identity affiliation annotation}}
We discern the social identity that users affiliate themselves with by comparing their usernames against a lexicon of identities from a 2015 population survey of US residents that consolidates representations of their occupation and affiliations \cite{smith2016mean}. The users are annotated with their identities, which are manually analyzed in the context of the \texttt{User} graph, and the manual unification of identity sets are annotated on the network graph itself.
These identities are classified into 7 categories (political, family, gender, race/nationality, religion, job, other), and the proportion of the categories are compared between bridging and non-bridging users.

\section{Results}
\begin{figure*}
    \centering
    \includegraphics[width=\linewidth]{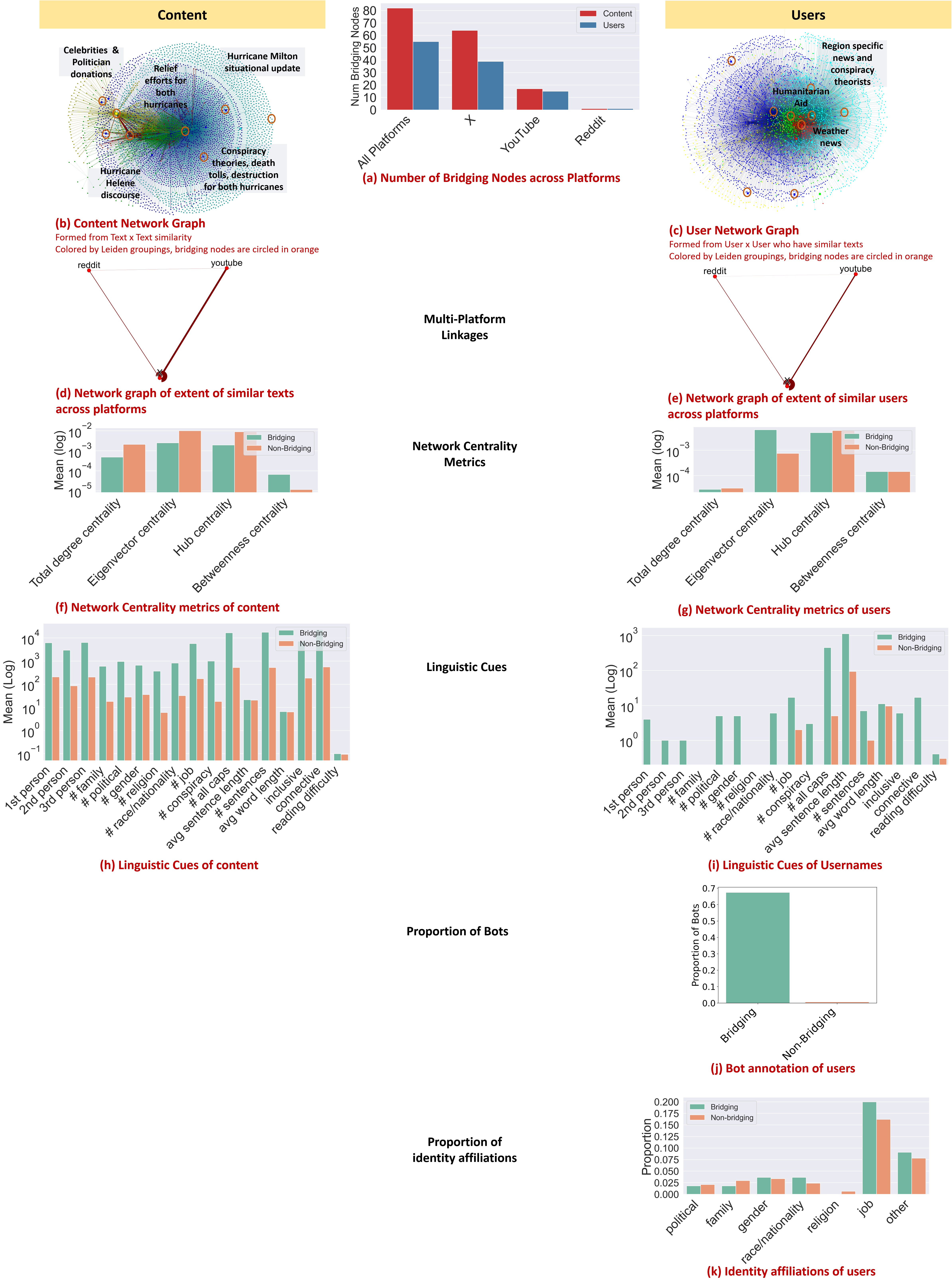}
    \caption{Comparison of bridging and non-bridging nodes}
    \label{fig:metrics}
\end{figure*}

\autoref{fig:metrics} presents the results from our analyses. The \texttt{Content} graph has 8,080 links, while the \texttt{User} graph has 35,889 links. Despite the extensive linkages (representing the similarity of nodes to each other) within the networks, the proportion of bridging nodes across all platforms are small ($<1\%$, \autoref{fig:metrics}(a)). Across both Content and Users, X hosts the highest number of bridging nodes, followed by YouTube and Reddit. Reddit is not an ideal platform for bridging nodes as its communities are segregated into distinct subreddits. 17.9\% of the links in the \texttt{User} graph and 24.2\% of links in the \texttt{Content} are cross-platform links (\autoref{fig:metrics}(d)(e)). The most commonly linked \texttt{Content} and \texttt{User} is between X and YouTube. 

By definition of a bridge, the betweenness centrality of bridging nodes, whether content or user nodes, are higher than non-bridging nodes, connecting disparate parts of the network (\autoref{fig:metrics}(f)(g)). However, the eigenvector centrality metric behaves differently. Eigenvector centrality is higher for bridging users than non-bridging users, but lower for bridging content than non-bridging content.
This means that bridging users cluster around other influential users, while this clustering pattern does not hold for bridging content. These dynamics highlight the nuanced roles of bridging nodes in facilitating information flow.

Specific to each hurricane, the topics are the donations and situational updates ( \autoref{fig:metrics}(b)). However, bridging content comes in the form of donations, relief efforts and conspiracy theories that are related to both hurricanes. Such examples are: announcements of government funding for both hurricanes, or theories that the hurricanes are man-made because they are in close succession with each other. In particular, bridging content connect groups of content that are talking about one hurricane, and morph the content to talking about the commonalities between both hurricanes. This therefore informs the types of users in the \texttt{User} graph, where the most common user affiliations are news broadcasters, humanitarian aid organizations, region specific news agencies and conspiracy theorists (\autoref{fig:metrics}(c)). These users typically want to reach multiple user communities to broadcast their message, be it a situational update, disaster aid information or conspiracy. Therefore, their bridging property serves to enhance their purpose in the graph.

Bridging content is typically complex. This is observed in \autoref{fig:metrics}(h)(i), where the mean of the linguistic cues for bridging content is higher than that of non-bridging content. The complexity of bridging content thus allows it to cover a breadth of topics and be similar to multiple sets of other content, as compared to the simpler non-bridging content. Bridging users have usernames that are longer (average sentence length) and have words in capital letters (\# all caps), perhaps to attract attention to their content.

From the bot annotation (\autoref{fig:metrics}(j)), 67.2\% of the bridging users are bots, while 0.5\% of the non-bridging users are bots. Bridging users straddle between multiple communities of users, which breaks the principle of homophily, where users congregate together because of common interests \cite{mcpherson2001birds}. It is easier for bots, who are inorganic users controlled by a computer program, to make connections with multiple groups. Such heterophily is harder for humans to maintain, with studies showing that heterophily can result in poor physical and mental health \cite{kim2023network}.

In terms of social identity, 34.5\% of bridging users and 27.8\% of non-bridging users present an identifiable social identity (\autoref{fig:metrics}(k)). Most of the identities extracted are job-related, which may be an artifact of the identity survey. Bridging users affiliate more with gender, race, nationality and job identities than non-bridging users. Such identities are common identities that all humans can relate to, as compared to specific interests, which suggests the effectiveness of bridging users to lean towards these identities. 

\section{What makes an Effective Bridge?}
When a bridging node is removed from the graph, the number of connected components in the graph increases. That is, if there were originally 1 connected component in the graph, a bridge removal results in the graph having 2 or more connected components. Since we used that algorithm to identify bridges, the bridges in our study, are by definition, ``effective", and produced the desired result in fragmenting the graph. However, fragmenting an extremely interconnected graph such as a social media discourse is a complex operation, therefore, we do not have many bridges.

Our methodology allows establishment of general characteristics of social media bridges:
\begin{itemize}
    \item Platform: Bridges tend to be nodes on X. X is a platform that connects multiple platforms via summaries or URLs (to the Reddit discussion thread or YouTube video) \cite{murdock2023identifying}, and are therefore effective bridges. The feed algorithm on X is created to maintain a balance and diverse feed, which allows crossover of content types \footnote{\url{https://blog.x.com/engineering/en_us/topics/open-source/2023/twitter-recommendation-algorithm}}. In contrast, Reddit gatekeeps users by interest-based subreddits, and YouTube where its recommendation algorithm is based on the user's past watched videos.
    \item Content: Bridging content are complex, having long sentences, many inclusive and connective words, and many pronouns. Such complex content are effective bridges since they can bridge multiple communities.
    \item User type: Bots are effective bridging users for they can easily straddle different sets of user communities and take on multiple personas, even changing affiliations. Identity affiliations related to gender, race and job are commonly used among bridging users compared to interest-specific affiliations. Influential users (high eigenvector centrality) are more primed to be bridging users.
\end{itemize}


Understanding social media bridges can significantly inform disaster management strategies. During natural disasters, governments and non-profit organizations aim to disseminate situational updates, aid-related information, and combat misinformation \cite{zhang2019social}. To maximize the reach and acceptance of their messages, these organizations' social media personas should resonate with diverse communities. By positioning these personas as bridges, organizations can leverage various strategies, such as adopting characteristics of effective bridges, natural identity affiliations, or influential users (e.g., politicians or local figures). Automation in the form of bots can be used for wider message reach, though malicious bots can also be used to propagate misinformation and conspiracy theories during natural disasters \cite{phillips2022hoaxes}. Messages can also be crafted with complex sentence structures that address multiple topics simultaneously. Furthermore, platforms like X often offer strategic advantages for broad dissemination of disaster-related information.

\paragraph{\textbf{Dataset Availability}} We release our dataset at \url{https://doi.org/10.5281/zenodo.14791336}. The dataset includes text content, user details, and annotations of the bridging properties of both content and users. 

\paragraph{\textbf{Limitations and Future Work}} Our dataset has some limitations due to restricted access to APIs, which may result in an incomplete view of the discourse. However, data from multiple platforms were collected to provide a more comprehensive perspective. Additionally, as noted in \cite{ng2023recruitment}, the list of social identities used may not fully align with the descriptions people use on social media, necessitating regular updates to stay current with evolving social media language. Future work could explore the evolution of content through bridges to gain deeper insights into their functionality.

\section{Conclusion}
Our study introduces the concept of bridging nodes on social media. Our investigations trace social media discourse on two 2024 US hurricanes on X, YouTube and Reddit. By modeling the similarity of narratives through network graphs, we can identify how these narratives are connected (via content), and who connect narratives (via users). These connections are termed as \emph{bridging nodes}. While only less than 1\% of nodes are bridging nodes, these nodes serve as important links for information flow. Removal of these bridges breaks up the the connectivity of the graph, and therefore the path of information transversal breaks too. Deeper analysis of these bridging nodes through network cues, linguistic cues, topic analysis and user information show key properties that differentiate these bridging nodes from non-bridging nodes. We hope our work facilitates downstream studies of information diffusion, and can inform public officials and disaster aid organizations on how they should position themselves and their content for announcements to reach diverse communities.


\bibliographystyle{ACM-Reference-Format}
\bibliography{references}



\end{document}